\documentclass[structabstract]{aa} 
\usepackage{graphicx} 
\usepackage{txfonts} 
\usepackage{natbib} 
\bibpunct{(}{)}{;}{a}{}{,} 
\begin{document}

\title{SDSS surface photometry of M\,31 with absorption corrections}

\titlerunning{SDSS photometry of M31}

\author{E.~Tempel\inst{1,}\inst{2} \and T.~Tuvikene\inst{1,}\inst{3} \and A.~Tamm\inst{1} \and P.~Tenjes\inst{1,}\inst{2}}

\institute{Tartu Observatory, Observatooriumi~1, 61602 T\~oravere, Estonia \\
\email{[elmo;taavi;atamm]@aai.ee} \and Institute of Physics, Tartu University, T\"ahe~4, 51010 Tartu, Estonia \\
\email{peeter.tenjes@ut.ee} \and Vrije Universiteit Brussel, Pleinlaan 2, 1050 Brussels, Belgium}

\date{Received 04 November 2010 / Accepted 09 December 2010}

\abstract
{} 
{The objective of this work is to obtain an extinction-corrected distribution of optical surface brightness and colour indices of the large nearby galaxy M\,31 using homogeneous observational data and a model for intrinsic extinction.} 
{We process the Sloan Digital Sky Survey (SDSS) images in $ugriz$ passbands and construct corresponding mosaic images, taking special care of subtracting the varying sky background. We apply the galactic model developed in Tempel et al. (2010) and far-infrared imaging to correct the photometry for intrinsic dust effects. } 
{We obtain observed and dust-corrected distributions of the surface brightness of M\,31 and a map of line-of-sight extinctions inside the galaxy. Our extinction model suggests that either M\,31 is intrinsically non-symmetric along the minor axis or the dust properties differ from those of the Milky~Way. Assuming the latter case, we present the surface brightness distributions and integral photometry for the Sloan filters as well as the standard $U\!BV\!RI$ system. We find the following intrinsic integral colour indices for M\,31: $(U-B)_0=0.35$; $(B-V)_0=0.86$; $(V-R)_0=0.63$; $(R-I)_0=0.53$; the total intrinsic absorption-corrected luminosities of M\,31 in the $B$ and the $V$ filters are 4.10 and 3.24\,\mbox{mag}, respectively. } 
{}

\maketitle

\section{Introduction} 

\begin{figure}	
    \centering
    \includegraphics[width=73mm]{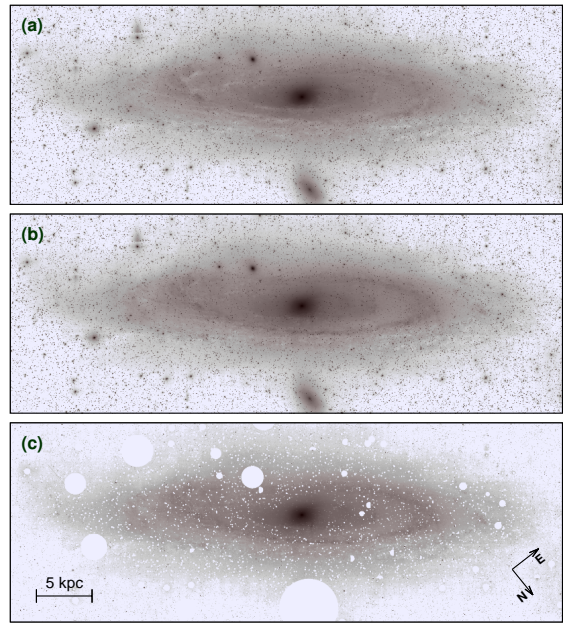}
    \caption{Upper panel~(a): SDSS imaging of M\,31 in $g$ filter, corrected for background variations. Middle panel~(b): the same as upper panel, with intrinsic extinction effects removed (see Fig.~\ref{fig:tau}b for extinction map). Lower panel~(c): the same as middle panel, showing also the masked regions (white circles).} 
    \label{fig:m31_pix} 
\end{figure}
\begin{figure*}
    \centering
    \includegraphics[width=175mm]{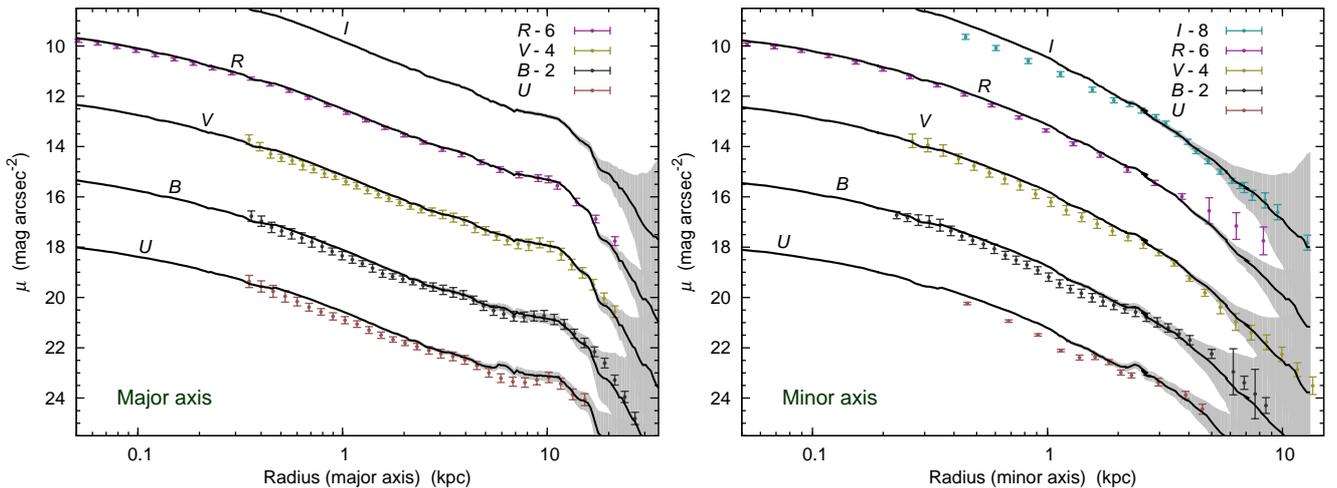} 
    \caption{Elliptically averaged observed surface brightness profiles from the SDSS photometry, converted to the $U\!BV\!RI$ system (black solid lines) together with uncertainties (grey regions). Conversion to the $U\!BV\!RI$ system is done according to \citet{Blanton:07}. For comparison, some earlier measurements are also shown (points with error bars; for references, see Sect.~\ref{results}). The earlier measurements in $U\!BV\!R$ are from elliptically averaged data, while $I$ and outermost parts of $V$ measurements by \citet{Irwin:05} have been made along the minor axis only. Left panel: major axis; right panel: minor axis. For clarity, the $BV\!RI$ profiles have been shifted by the indicated values. } 
    \label{fig:obs_ubvri} 
\end{figure*}

The primary information about the structure of galaxies and their stellar populations comes from the distribution of surface brightnesses and colour indices. Thus, it is only natural that the photometry of the largest nearby galaxy M\,31 has been studied thoroughly for almost a century \citep[for references, see][]{van-den-Bergh:91,Tenjes:94,Tempel:10}. More recently, satellite observations in the far-infrared have become available and attempts have been made to consider the impact of dust extinction on the observable properties of M\,31 \citep{Xu:96,Haas:98,Montalto:09,Tempel:10}. While the general luminosity distributions of different studies are in good agreement, the distribution of colours is not so well settled. Small systematic deviations between different datasets lead to considerable uncertainties of colour indices -- too big for studying in detail the stellar populations or the star formation history of the galaxy with the help of chemical evolution models.

Recently, within the Sloan Digital Sky Survey \citep[SDSS; ][]{York:00}, a contiguous strip covering the entirety of M\,31 through $ugriz$ filters has been observed. The combination of five filters on a single, well-calibrated telescope and roughly uniform atmospheric conditions during the observations provide a uniquely homogeneous dataset for a thorough analysis of the extensive galaxy. In this paper, we use the SDSS observations to study the detailed luminosity and colour distribution of M\,31. We use also far-infrared imaging by the Spitzer Space Telescope and Infrared Astronomical Satellite to correct the derived photometry from dust extinction.

Throughout the paper, we assume the distance of M\,31 to be 785\,\mbox{kpc} \citep{McConnachie:05}, corresponding to the scale \hbox{1$\arcmin$ = 228\,\mbox{pc}}. The inclination angle of the galaxy has been taken 77\fdg5 \citep{Walterbos:88,deVaucouleurs:91}. According to \citet{Walterbos:87}, \citet{Ferguson:02}, and our analysis of the SDSS and Spitzer images, we set 38\fdg1 as the major axis position angle. All luminosities and colour indices have been corrected from extinction in the Milky~Way according to \citet{Schlegel:98} and are given in AB-magnitudes for the $ugriz$ filters and in Vega magnitudes for the $U\!BV\!RI$ filters, as usual. For the Sloan filters, the extinction is derived from the \citet{Schlegel:98} estimates and the Galactic extinction law by linear interpolation. The absolute solar luminosity for each filter was taken from \citet{Blanton:07}. The applied solar luminosities and the Galactic extinction for each filter are presented in Table~\ref{table:filters}. 
\begin{table}
    \caption{The applied solar luminosities (in AB~magnitudes for $ugriz$ filters and in Vega magnitudes for $U\!BV\!RI$ filters) and Galactic extinctions~($A$) for each filter.} 
    \label{table:filters} 
    \centering 
    \begin{tabular}
        {llllll} 
        \hline\hline 
        & $u$ & $g$ & $r$ & $i$ & $z$ \\
        \hline 
        M$_{\sun}$ & 6.38 & 5.12 & 4.64 & 4.53 & 4.51 \\
        $A$ & 0.320 & 0.250 & 0.175 & 0.135 & 0.095 \\
        \hline 
        & $U$ & $B$ & $V$ & $R$ & $I$ \\
        \hline
     M$_{\sun}$ & 5.55 & 5.45 & 4.78 & 4.41 & 4.07 \\
        $A$ & 0.337 & 0.268 & 0.206 & 0.166 & 0.120 \\
        \hline 
    \end{tabular}
\end{table}

\section{SDSS data reduction}

M\,31 was observed with the Sloan telescope on October~6, 2002, using $ugriz$ filters. We retrieved the pipeline-corrected frames (fpC-files) from the SDSS Data Archive Server (DAS). The frames (2048$\times$1489 pixels, 0.396\,arcsec\,px$^{-1}$) originate from two observing runs obtained with six columns of CCDs, thus giving a total of 12 scanlines parallel to the apparent major axis of the galaxy. The scanlines have a width of 13.52~arcmin, overlapping by about 55~arcsec. We used 56 frames from each scanline (field numbers 49--104), in order to construct mosaic images in $ugriz$ passbands.

Individual frames were aligned by applying astrometric transformations given in frame headers. In the final mosaics, 10$\times$10 binning was implemented, resulting in image dimensions of 2300$\times$7400 pixels. For the central region of M\,31, additional images with higher sampling were created using 2$\times$2 binning.

Sky background was estimated by analysing the SDSS frames in several steps. We used the \emph{sky.pro} routine from the IDL Astronomy User's Library\footnote{http://idlastro.gsfc.nasa.gov/} which gives an estimated mode of the sky-background values. The total sky contribution in each pixel, $sky_\mathrm{total}$, was assumed to consist of three components:
\begin{equation}
    sky_\mathrm{total} = sky_\mathrm{scanline} + sky_\mathrm{temporal} + sky_\mathrm{gradient}. 
\end{equation}
The base sky value, $sky_\mathrm{scanline}$, was determined at both ends of each scanline, at a sufficient distance from M\,31. Superimposed on it, variations along and perpendicular to the scanlines were found, probably caused by different effects. Most likely, the former is the result of temporal sky brightness variations during the observations and is thus designated here as $sky_\mathrm{temporal}$. It was determined on the basis of the two outermost scanlines, separately for both observing runs, with a moving window with the width of a full frame and the height of 1000 pixels, yielding a smooth curve along the scanline. Sky variations perpendicular to the scanlines (and perpendicular to the major axis of M\,31), $sky_\mathrm{gradient}$, were probably caused by stray light from the galaxy itself. Its contribution was determined by matching the sky-subtracted signal in overlapping regions of side-by-side scanlines. The values of $sky_\mathrm{gradient}$ were then linearly interpolated between the start and the end of each pixel row of each scanline.

After subtracting the soft bias of 1000\,ADU (added in the SDSS pipeline) and the sky background, the pixel values were corrected for atmospheric extinction and were converted to the flux and magnitude scale, following the SDSS photometric flux calibration\footnote{http://www.sdss.org/dr7/algorithms/fluxcal.html\#counts2mag}. In order to eliminate the contribution of foreground stars and the galaxies M\,32 and NGC\,205, circular masks were applied on top of these objects (see Fig.~\ref{fig:m31_pix}c). SExtractor \citep{Bertin:96} was used to detect and mask faint background galaxies in the outer regions.

For the uncertainty estimation, the following noise sources were taken into account: the random photon noise, the contribution of dark current and readout noise, the uncertainty of the sky level estimate (0.5\,ADU), and the error of matching brightness levels in the overlapping regions (0.5\,ADU). More details about the conducted image processing and sky removal will be provided in a separate paper (in preparation).

The final image of M\,31 for the $g$ filter is shown in Fig.~\ref{fig:m31_pix}a. The applied masks are shown in Fig.~\ref{fig:m31_pix}c. 
\begin{figure}
    \centering
    \includegraphics[width=88mm]{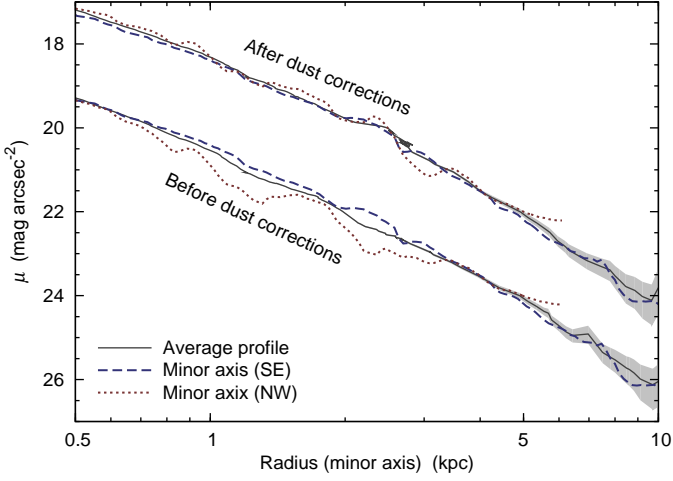} 
    \caption{Surface brightness profiles along the minor axis before and after intrinsic dust corrections. Red dotted line shows the profile along the nearer side (NW direction); blue dashed line shows the profile along the further side (SE direction). The corresponding elliptically averaged surface brightness profiles are also shown (black solid lines) together with error estimates (grey region). For clarity, the dust-corrected profiles have been shifted up by two units.} 
    \label{fig:profiles} 
\end{figure}

\section{Structural model and extinction corrections}\label{mod} 
\begin{figure}
    \centering
    \includegraphics[width=88mm]{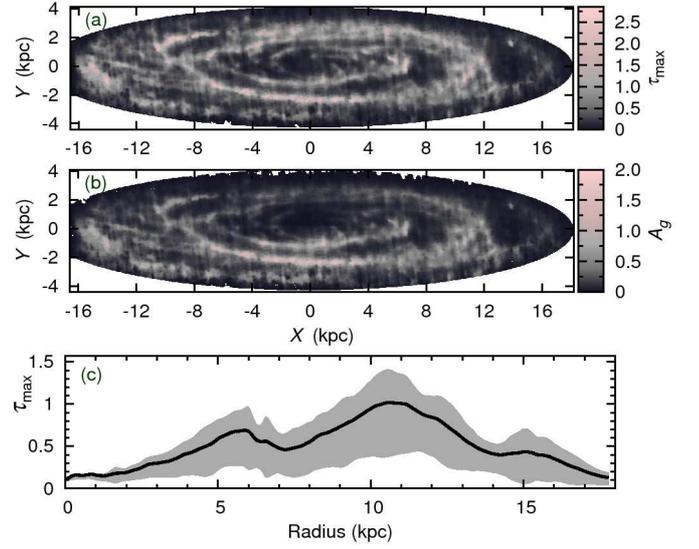}
    \caption{Panel~(a): map of the total optical depth $\tau_{\mathrm{max}}$ for the Sloan $g$ filter; panel~(b): extinction map for the $g$ filter, calculated according to the best-fit model; panel~(c): the elliptically averaged profile of the optical depth map for the $g$ filter; grey region indicates the scatter of the optical depth along each ellipse.} 
    \label{fig:tau} 
\end{figure}

To correct the optical images for dust extinction, we have to consider the spatial distribution of stars and dust inside the galaxy. For this purpose, we have relied on a general three-dimensional galaxy model and extinction calculation methods, briefly reviewed below. More details can be found in \citet{Tempel:10}.

The model galaxy is given as a superposition of its individual stellar components and a dust disc. Each stellar component is approximated by an ellipsoid of rotational symmetry with a constant axial ratio $q$; its spatial density distribution follows Einasto's law 
\begin{equation}
    \label{eq:explaw} l(a)=l_0\exp{\left[-\left(\frac{a}{ka_0}\right)^{1/N}\right]}, 
\end{equation}
where $l_0=hL/(4\pi q a_0^3)$ is the central density and $L$ is the component luminosity; $a=\sqrt{r^2+z^2/q^2}$, where $r$ and $z$ are two cylindrical coordinates; $a_0$ is the harmonic mean radius. The coefficients $h$ and $k$ are normalising parameters, depending on \emph{N}.

In the case of the young disc the spatial density distribution is assumed to have a toroidal form, approximated as a superposition of a positive and a negative density component, both following Eq.~\ref{eq:explaw}.

Density distributions of all stellar components are projected along the line of sight and their sum yields the surface brightness distribution of the model 
\begin{equation}
    L(X,Y)=\sum\limits_{k}\int\limits^{\infty}_{X} \frac{\sum\limits_{j=1}^2\left[l_k(r,z_j)\,\mathrm{e}^{-\tau_k(r,z_j)}\right]} {\sin{i}\,\sqrt{r^2-X^2}}r\,\mathrm{d}r,\label{eq:dust_lumin} 
\end{equation}
\begin{equation}
    z_{1,2}= \frac{Y}{\sin{i}} \pm \frac{\sqrt{r^2-X^2}}{\tan{i}}, 
\end{equation}
where $l(r,z)$ denotes the spatial luminosity density (Eq.~\ref{eq:explaw}) of the galaxy component, $L(X,Y)$ is the corresponding surface brightness distribution, $X$ and $Y$ are coordinates in the plane of the sky, $\tau(r,z)$ is the optical depth of dust along the line of sight for the given location, $i$ is the angle between the sky plane and the galaxy plane, and the summation index $k$ designates each visible component.

A dust disc is added to the model galaxy, with its vertical density distribution following Eq.~\ref{eq:explaw}. Along each line of sight, the optical depth $\tau(r,z)$ is assumed to be proportional to the dust column density. The optical depth is taken zero between the observer and the dust disc, and equal to $\tau_{\mathrm{max}}(X,Y)$ behind the dust disc. $\tau_{\mathrm{max}}$ is the total optical depth along a given line of sight and is thus a function of $(X,Y)$. Inside the dust disc, $\tau(r,z)$ is between $0$ and $\tau_{\mathrm{max}}$; see fig.~1 and eq.~6 in ~\citet{Tempel:10}.

Assuming that dust column density is proportional to the far-infrared flux, the map of $\tau_{\mathrm{max}}(X,Y)$ can be derived from far-infrared imaging: 
\begin{equation}
    \label{eq:dxy} \tau_{\mathrm{max}, f}(X,Y)=c_{f,\lambda',T'}F_{\lambda'}(X,Y) \frac{\lambda ^\beta B(\lambda',T'\,)} {(\lambda ')^\beta B(\lambda ,T(X,Y))}, 
\end{equation}
where $F_{\lambda'}(X,Y)$ is the far-infrared flux map at a reference wavelength $\lambda '$; $T '$ is a reference temperature; $T(X,Y)$ is a map of the dust temperatures; $B(\lambda,T)$ is the black-body function; $\beta$ is the dust emissivity index; and $c$ is an empirical calibration constant, corresponding to the reference wavelength and temperature, and to the filter $f$ of the optical observations ($c\propto A_{\lambda}/E(B-V)=R_VA_{\lambda}/A_V$).

We used the Spitzer Space Telescope observations at 24, 70 and 160\,$\mu$m and the Infrared Astronomical Satellite detections at 100\,$\mu$m to derive maps for the far-infrared flux and dust temperature. The point-spread-functions and sampling of the Spitzer observations were reduced to that of 160\,$\mu$m imaging. A dust model, comprising a colder and a warmer component was used to approximate the measured far-infrared spectral energy distribution along each line of sight with modified Planck functions; extinction was ascribed to the colder component only. More details about this step and the related error estimates can be found in \citet{Tempel:10}.

For applying Eq.~\ref{eq:dxy}, we chose $\lambda'=100\,\mu$m and $T'=18.2$\,K as in \citet{Schlegel:98}, and according to \citet{Vlahakis:05} we take the dust emissivity index $\beta=2.0$. The best value for the calibration constant $c$ was found during the fitting process. Because of the specific geometry and viewing angle of M\,31, the surface brightness distribution along the minor axis is sensitive to the optical depth. Extinction is higher along the closer side of the minor axis because most of the bulge stars are behind the dust disc, whereas for the farther side, the bulge is in front of the dust disc \citep[see fig.~1 in][]{Tempel:10}. However, assuming the galaxy to be symmetric, dust-free luminosity distributions along the minor axis in the two opposite directions should overlap, providing a recipe for the calibration of the constant $c$. During the iterative fitting process (see Sect.~\ref{results} and Fig.~\ref{fig:profiles}), the best overlap was reached with $c=0.022$ for the $g$ passband. The Galactic extinction law was used to calculate optical depths for observations with other filters.

\section{Results: model fitting and surface brightness profiles}
\label{results}

\begin{figure}
    \centering
    \includegraphics[width=78mm]{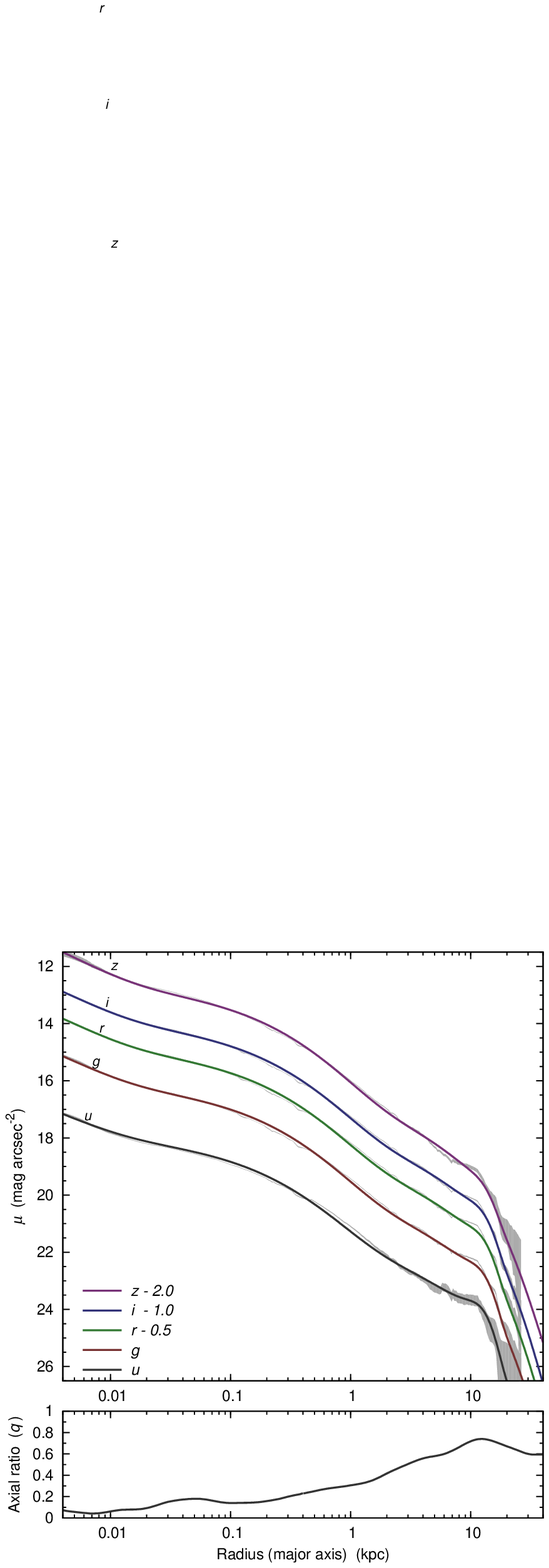} 
    \caption{Top panel shows the final model (solid lines) and observed (grey regions) profiles along the major axis in $ugriz$ filters. For clarity, the $r,i,z$ profiles have been shifted by the indicated values. The errors for the model are mainly due to the errors of the observations.  Bottom panel shows the observed axial ratio~$q$. The errors from the ellipse fitting for axial ratio $q$ are in order of 0.01.}
    \label{fig:model_prof_filt} 
\end{figure}

On the basis of the SDSS observations, we have derived the surface brightness distributions of M\,31 in $ugriz$ filters. Figure~\ref{fig:obs_ubvri} presents the elliptically averaged observational surface brightness profiles with error estimates in $U\!BV\!RI$ filters. The elliptically averaged profiles were generated using the IRAF/STSDAS task \emph{ellipse}. The conversions from the Sloan filters to the $U\!BV\!RI$ system were conducted according to \citet{Blanton:07}. For comparison, some earlier measurements from the literature are also shown (data points with error bars): \citet{de-Vaucouleurs:58} ($B$), \citet{Hoessel:80} ($U\!BV$), \citet{Hodge:82} ($B$), \citet{Kent:87} ($R$), \citet{Walterbos:88} ($U\!BV\!R$), \citet{Pritchet:94} ($V$), and \citet{Irwin:05} ($V\!I$). The $U\!BV\!R$ profiles are elliptically averaged profiles, the $I$ profile is measured along the minor axis of M\,31.

However, for applications in stellar population analyses, surface brightness and colour distributions corrected for intrinsic dust extinction are more desirable. The amount of extinction depends on the properties of the dust disc and on the general three-dimensional luminosity distribution of a galaxy. The corresponding model is briefly described in Sect.~\ref{mod}.

To fit the model to the observations, we have assumed the galaxy M\,31 to consist of a bulge, an old and a young disc, an outer halo, and a central nucleus. Model construction details are similar to the modelling presented in \citet{Tempel:10}.

For model fitting we used the Sloan $ugriz$ profiles. Model fitting was started with crude initial model parameters. In the first step of the fitting process, a stellar model was fitted to the elliptically averaged surface brightness profiles (using \emph{ellipse}), neglecting the dust disc and thus any intrinsic extinction. Next, the dust disc was added and the extinction map was calculated according to the optical depth map and the luminosity density distribution model, derived in the first step. Now the extinction map was applied to the full 2-dimensional optical imaging, yielding a dust-free image of the galaxy. The calibration constant $c$ in Eq.~\ref{eq:dxy} was determined, providing the best overlap of the dust-free surface brightness profiles along both sides of the minor axis. In Fig.~\ref{fig:profiles} the profiles along both sides of the minor axis before and after dust corrections are shown. Finally, the structural model was fitted to the extinction-corrected imaging, allowing us to constrain the model parameters. All these steps were repeated iteratively: according to the new model, a new extinction map was calculated and applied to the initial imaging; calibration constant $c$ was adjusted and a new extinction-corrected image was calculated, allowing us to further constrain the model parameters. Four iterations were needed to reach convergence. The final model is presented in Appendix~\ref{sect_1}.

Figure~\ref{fig:tau} shows the total optical depth map (panel a) and the corresponding extinction map (panel b) of M\,31 for the Sloan $g$ filter. Extinction is higher along the closer side of the minor axis due to the geometry and viewing angle of the galaxy. This asymmetry was used to determine the calibration constant $c$ in Eq.~\ref{eq:dxy} during the fitting process as described above. Figure~\ref{fig:tau}c presents the elliptically averaged profile of the total optical depth for the $g$ filter; the grey region illustrates deviations of the optical depth values along each ellipse. In Fig.~\ref{fig:tau}c, the signature of three dust rings can be seen.

\begin{table}
    \caption{Visible and intrinsic colour indices of M\,31, corrected for extinction in the Milky~Way.} 
    \label{table:colors} 
    \centering 
    \begin{tabular}
        {lllll|c} 
        \hline\hline 
        M\,31& $u-g$ & $g-r$ & $r-i$ & $i-z$ & $g$ \\
        \hline 
        Intrinsic & 1.58 & 0.75 & 0.40 & 0.21 & 3.71 \\
        Visible$^{(a)}$& 1.68 & 0.80 & 0.43 & 0.24 & 3.86 \\
        \hline 
        M\,31& $U-B$ & $B-V$ & $V-R$ & $R-I$ & $B$ \\
        \hline 
        Intrinsic & 0.35 & 0.86 & 0.63 & 0.53 & 4.10 \\
        Visible$^{(b)}$& 0.43 & 0.90 & 0.65 & 0.58 & 4.27 \\
        de Vaucouleurs & 0.43 & 0.86 & & & 4.36 \\
        \hline 
    \end{tabular}
    \begin{list}
        {}{} 
        \item[] Notes: $^{(a)}$ estimated errors are 0.09\,mag for $u-g$ and 0.06\,mag for other colour indices. $^{(b)}$ estimated errors are on the order of 0.10\,mag. Colour indices given by \citet{deVaucouleurs:91} are visible colour indices. 
    \end{list}
\end{table}

Figure~\ref{fig:m31_pix} displays the background-subtracted $g$-filter mosaic images of M\,31 before (panel a) and after (panel b) dust extinction corrections. The profiles presented in Fig.~\ref{fig:profiles} are measured from these images.

The derived intrinsic $ugriz$ surface brightness profiles along the major axis of the best model are shown in Fig.~\ref{fig:model_prof_filt}. The lower panel of Fig.~\ref{fig:model_prof_filt} shows the axial ratio $q$ of elliptically averaged profiles of M\,31.

\begin{figure}
    \centering
    \includegraphics[width=80mm]{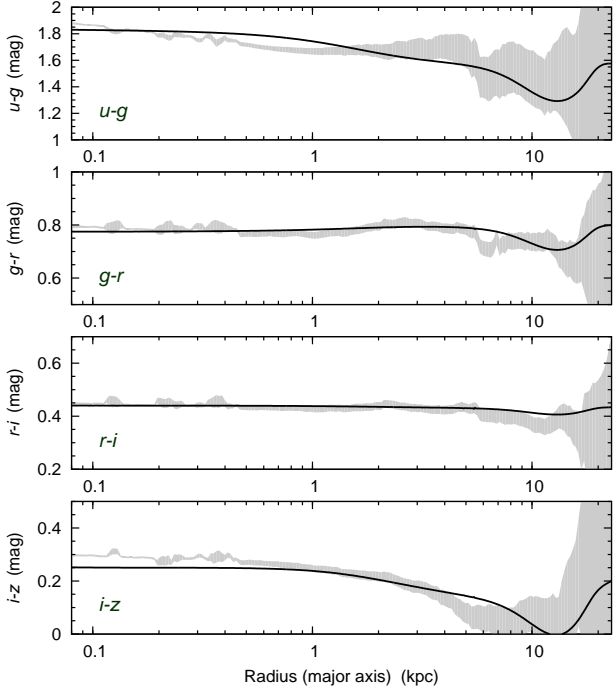} 
    \caption{Distribution of the elliptically averaged observed (grey regions) and modelled (solid lines) colour indices for the Sloan filters along the major semi axis. The scatter of the grey region indicates the uncertainty of observations and ellipse fitting. The colour indices are corrected for the absorption in the Milky~Way and for the intrinsic absorption of M\,31.}
    \label{fig:colors_sdss} 
\end{figure}

\begin{figure}
    \centering
    \includegraphics[width=80mm]{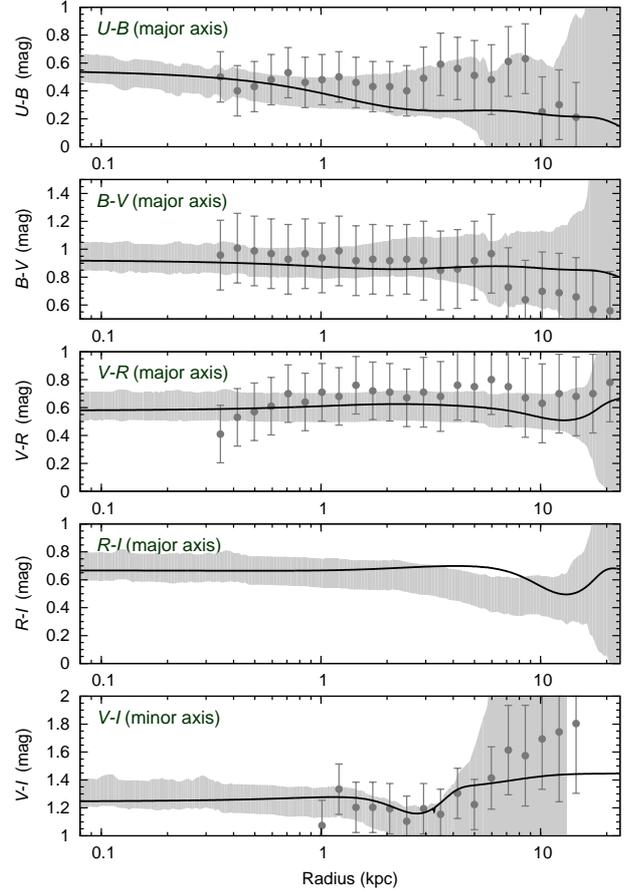} 
    \caption{Distribution of the elliptically averaged observed (grey regions) and modelled (solid lines) colour indices for the $U\!BV\!RI$ filters along the major (four top panels) and minor (bottom panel) semi axis. The scatter of the grey region indicates the uncertainty of observations, ellipse fitting and conversion to the $U\!BV\!RI$ system. Points with error-bars are observational data as compiled in \citet{Tempel:10}, shown for comparison; the $I$ data were not available for the major axis.} 
    \label{fig:colors_ubvri} 
\end{figure}

In Fig.~\ref{fig:colors_sdss}, the distributions of the colour indices of the model (solid lines) are compared to the observations (grey regions) for the Sloan filters. In Fig.~\ref{fig:colors_ubvri}, the same is shown for the standard $U\!BV\!RI$ filters. In the latter case, the data points with error bars are observational data from the literature as compiled in \citet{Tempel:10}. The colours of M\,31 are given in Table~\ref{table:colors}. The colours are corrected for the extinction in the Milky~Way. For comparison, the colour indices derived by \citet{deVaucouleurs:91} are given. 

 
The derived value for the constant $c=0.022$ can be compared to the constant $p$ in eq.~22 of \citet{Schlegel:98} for the Milky~Way. In case of the Galactic extinction law, $p=0.016$ would transform to $c= \ln(10) / 2.5 \times A_g/(A_B-A_V) \times p=0.056$ for the Sloan $g$ filter. Using this value in the case of M\,31, the restored intrinsic brightness along the minor axis on the side closer to us would be higher than on the farther side, because the closer side is more affected by dust absorption, as explained in the last paragraph of Sect.~\ref{mod}. Therefore, our model suggests that either  M\,31 is intrinsically non-symmetric, with the closer side being brighter than the farther side at respective distances from the centre, or the dust properties of M\,31 differ from those of the Milky~Way, leading to a different value of $c$. The former scenario is not supported by the Spitzer Space Telescope observations at $3.6\,\mu$m \citep{Barmby:06}, where the stellar light should be least affected by absorption. The latter scenario remains to be confirmed or invalidated with future observations of M\,31 with the Herschel Space Observatory, providing higher angular resolution in far-infrared than the Spitzer Space Telescope, and a coverage of longer wavelengths.

The derived luminosity profiles and the resulting distributions of colour indices are based on a homogeneous dataset. Therefore, possible systematic errors in the colour profiles are smaller than in profiles derived by averaging data from different instruments.  The improved spatial coverage and lower uncertainties provided by the SDSS data, demonstrated by Fig.~\ref{fig:colors_ubvri}, allow us to obtain tighter constraints for the star formation history and the overall origin of M31. 

\begin{acknowledgements}
    We thank the anonymous referee for useful comments, helping to improve the readability of the paper. We acknowledge the financial support from the Estonian Science Foundation {grants 7146, 7765, 7115} and the Estonian Ministry for Education and Science research project SF0060067s08. The authors thank Stephen Kent for useful comments on SDSS imaging. All the figures have been made with the gnuplot plotting utility.

    We are pleased to thank the SDSS Team for the publicly available data releases. Funding for the SDSS and SDSS-II has been provided by the Alfred P. Sloan Foundation, the Participating Institutions, the National Science Foundation, the U.S. Department of Energy, the National Aeronautics and Space Administration, the Japanese Monbukagakusho, the Max Planck Society, and the Higher Education Funding Council for England. The SDSS Web Site is http://www.sdss.org/.

    The SDSS is managed by the Astrophysical Research Consortium for the Participating Institutions. The Participating Institutions are the American Museum of Natural History, Astrophysical Institute Potsdam, University of Basel, University of Cambridge, Case Western Reserve University, University of Chicago, Drexel University, Fermilab, the Institute for Advanced Study, the Japan Participation Group, Johns Hopkins University, the Joint Institute for Nuclear Astrophysics, the Kavli Institute for Particle Astrophysics and Cosmology, the Korean Scientist Group, the Chinese Academy of Sciences (LAMOST), Los Alamos National Laboratory, the Max-Planck-Institute for Astronomy (MPIA), the Max-Planck-Institute for Astrophysics (MPA), New Mexico State University, Ohio State University, University of Pittsburgh, University of Portsmouth, Princeton University, the United States Naval Observatory, and the University of Washington. 
\end{acknowledgements}


\begin{appendix}
    
    \section{Structural model of M 31} \label{sect_1}

    The derived extinction distribution assumes a certain three-dimensional light distribution model, consistent with the observed surface brightness distributions along the major and minor axis of the galaxy.

    The final model parameters are given in Table~\ref{table:param}. These parameters correspond to the dust-free galaxy model, i.e. they represent the intrinsic stellar emission of the galaxy. The model given in Table~\ref{table:param} is not unique: the same total surface brightness profiles can be achieved using slightly different component parameters. To determine the real components of the galaxy, metallicity determinations and galaxy dynamics should be taken into account. For the present study, however, realistic deviations from the given model would cause negligible effects.

\begin{figure}
        \centering
        \includegraphics[width=78mm]{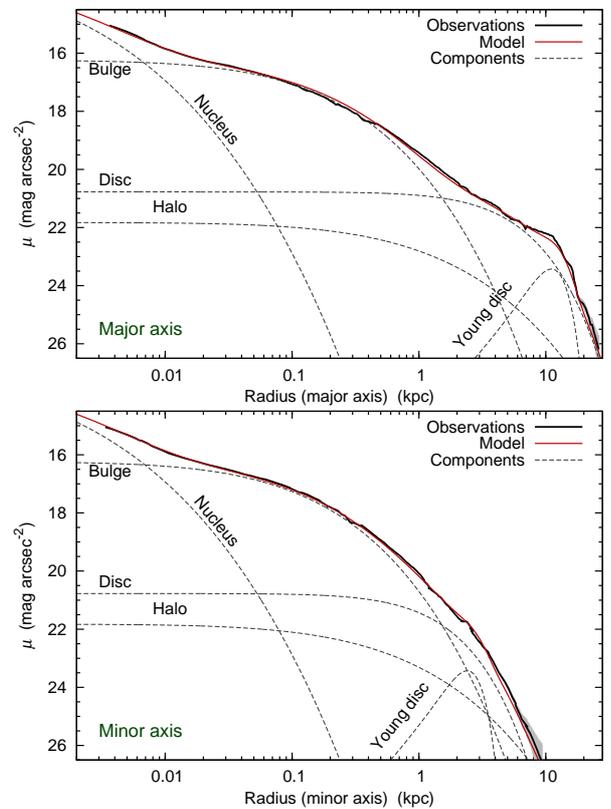} 
        \caption{Extinction-corrected surface brightness distribution of M\,31 along the major (upper panel) and minor (lower panel) axis in Sloan $g$ filter. Individual stellar components are given with dashed lines.} 
        \label{fig:model_profiles} 
    \end{figure}

    In Fig.~\ref{fig:model_profiles} the contributions by each galactic component to the surface brightness distribution in the $g$ filter are compared to the profile measured from the corresponding dust-corrected image. 
    \begin{table}
        \caption{Calculated parameters of the photometric model.} 
        \label{table:param} 
        \centering 
        \begin{tabular}
            {llllll} 
            \hline\hline 
            Population & $a_0$ & $q$ & $N$ & $k$ & $h$ \\
            & (kpc) & & & & \\
            \hline 
            Nucleus &0.01&1.00&4.0&$1.2626\cdot10^{-4}\!\!\!$&3111.43 \\
            Bulge &0.63&0.72&2.7&$7.2288\cdot10^{-3}\!\!\!$&158.920 \\
            Disc & 7.70&0.17&1.20&$3.3418\cdot10^{-1}\!\!\!$&6.0071 \\
            Young disc$^{(a)}\!\!\!$& 10.00 &0.01&0.2&1.4895& 1.0160 \\
            Halo & 6.30 &0.50&3.0 &$2.9762\cdot10^{-6}\!\!\!$&313.600 \\
            Dust disc & 15.0 &0.01 &1.0 &0.5000&4.00000 \\
            \hline 
            Population & $L_u$ & $L_g$& $L_r$ & $L_i$& $L_z$\\
            \hline 
            Nucleus & \small{0.0008}& \small{0.0018}& --& --& \small{0.0050} \\
            Bulge & 0.34& 0.58 & 0.76 & 1.03 & 1.29 \\
            Disc & 0.79& 1.07 & 1.43 & 1.92 & 2.05 \\
            Young disc$^{(b)}\!\!\!$& 0.29& 0.24 & 0.26 & 0.33 & 0.25 \\
            Halo & 0.17& 0.21 & 0.29 & 0.40 & 0.80 \\
            \hline 
        \end{tabular}
        \begin{list}
            {}{} 
            \item[] Notes:
            $^{(a)}$ Structural parameters for the young disc are given for the positive component only. $^{(b)}$ The luminosity of the young disc is its true luminosity ($L=L_{+}+L_{-}$, where $L_{-}=-0.5L_{+}$). The luminosities are corrected for the intrinsic extinction and the extinction in the Milky~Way, and are given in the units of $10^{10}\rm{L_{\sun}}$. 
        \end{list}
    \end{table}
\end{appendix}

\end{document}